\documentclass{elsart}
\usepackage{graphicx,amssymb}
\journal{Physics Letters A}
\usepackage[centertags]{amsmath}
\usepackage{epsfig}
\usepackage{amscd}
\usepackage{amscd}
\usepackage{color}

\newcommand{\beq}{\begin{equation}}
\newcommand{\eeq}{\end{equation}}
\newcommand{\ba}{\begin{array}}
\newcommand{\ea}{\end{array}}
\newcommand{\bea}{\begin{eqnarray}}
\newcommand{\eea}{\end{eqnarray}}
\newcommand{\dsfrac}[2]{\frac{\displaystyle #1}{\displaystyle #2}}

\begin{document}
\begin{frontmatter}

\title{Quantum discord versus second-order MQ NMR coherence intensity in dimers}
\author{E.I.Kuznetsova},
\ead{kuznets@icp.ac.ru}
\author{A.I.Zenchuk\corauthref{cor}}
\corauth[cor]{Corresponding author.}
\ead{zenchuk@itp.ac.ru}
\address{
Institute of Problems of  Chemical Physics RAS, Chernogolovka, Moscow Region, 142432, Russia}

%{\bf   ,,,, }
%\begin{titlepage}
%\begin{center}\date{}
%\title{Discord in biparticle spin system}
%\end{center}
%\end{titlepage}

\begin{abstract}
We consider the quantum discord in a two-particle spin-1/2 system (dimer) governed by the standard multiple quantum (MQ) Hamiltonian and reveal the relation between the discord and the intensity of the second-order MQ coherence. Since the coherence intensities  may be measured in experiment,  we obtain a method of detecting the quantum discord in experiment. Comparison of discord and concurrence is represented.
\end{abstract}


\end{frontmatter}
%\maketitle

\section{ Introduction}

For a long time the quantum entanglement \cite{Werner,HWootters,P,AFOV,DPF} has been considered as a proper measure of quantum correlations. However,  
it seemed out that entanglement does not cover all quantum correlations which are responsible for advantages of quantum processes in comparison with their classical counterparts.  For instance, the quantum non-locality without entanglement is studied in \cite{BDFMRSSW,HHOSSS,NC}, the quantum  speed-up with separable states is demonstrated in \cite{BCJLPS,M,DFC,BBKM,DV,DSC}. 

{ Quantum discord  was first introduced in ref.\cite{Z0} and then it was studied in refs.\cite{HV,OZ,Z} as an alternative measure of quantum correlations.} This measure involves  larger class of quantum correlation in comparison with entanglement. In particular,
it may be nonzero even for the states with  zero entanglement. { Therefore it was noted \cite{DSC} that namely discord might be responsible for advantages of quantum information devices, in particular, for  quantum speedup.
The concept of 
 quantum discord is intensively developing during last years. For instance, the operational interpretation of quantum discord in terms of
 state merging \cite{MD} and unitary invariant modification of discord \cite{Z_a} have been proposed. } However, only very special cases have been treated analytically  \cite{L,ARA,Xu}.  Moreover,  relation between  discord and  physical characteristics  of quantum system 
 is not well studied yet. 

This paper is devoted to the  investigation of discord in a system of two spin-1/2 particles  (dimer) on the preparation period of the MQ NMR experiment \cite{WSWP,BMGP} with the thermodynamic equilibrium initial state in the strong external magnetic field. 
We show  that the discord can be considered as a function of  the second-order coherence intensity similar to the entanglement \cite{FP}. This conclusion gives us  a method to measure the discord in the MQ NMR experiment. Thereby one should note ref.\cite{AFY} where the magnetic susceptibility has been introduced as a physical characteristic of a quantum system allowing one to measure the  entanglement in nitrosyl iron complexes, as well as 
ref. \cite{Yu} { where the relations between discord and  such physical characteristics as internal energy, specific heat,
 and magnetic susceptibility  have been studied.}

We derive the explicit formula for the discord in a dimer with the dipole-dipole interaction (DDI)  governed by the standard MQ NMR Hamiltonian \cite{BMGP} in Sec.\ref{Section:discord}. The dependence of  the discord on the second-order MQ NMR coherence intensity is studied in Sec.\ref{Section:coh}. The comparison with analogous dependence of the concurrence (as a measure of entanglement) is given in the same section. Conclusions are given in Sec.\ref{Section:conclusions}

%%%%%%%%%%%
\section{Discord in the system of two spin-1/2 particles with MQ Hamiltonian}
\label{Section:discord}
The dynamics of a system of two spin-1/2 particles with the DDI  on the preparation period of  the MQ NMR experiment \cite{WSWP,BMGP} in the strong external magnetic field is 
 governed by the standard MQ NMR Hamiltonian $H_{MQ}$, which reads 
\begin{equation}\label{HMQ}
H_{MQ}= \frac{D}{2}(I^+_1I^+_2+I^-_1I^-_2)
\end{equation}
in the rotating reference frame \cite{G},
where $D$ is the coupling constant between two spins, which reads  $\displaystyle D=\frac{\gamma\hbar}{ r^3_{12}}(1-3\cos^2 \theta_{12})$ in the case of  DDI. Here $\gamma$ is the gyromagnetic ratio, $r_{12}$ is the distance between spins,  $\theta_{12}$ is the angle between  the vector $\vec{r}_{12}$  and the external magnetic field $\vec{H}_0$; $I_j^+=I_{jx} + i I_{jy}$ and $I_j^-=I_{jx} - i I_{jy}$ ($j=1,2$) are the raising and lowering operators of spin $j$; $I_{\alpha,j}$ are operators of  the $j$th 
spin angular momentum projection on the axis $\alpha=x,y,z$. 
Starting with  the thermodynamic equilibrium initial state of the dimer in the strong external magnetic field,
\begin{eqnarray}\label{rho0}
\rho_0=\frac{e^{\beta I_z}}{{\mbox{Tr}} \left(
e^{\beta I_z}\right)},
\end{eqnarray}
({ where $I_z=I_{1z}+I_{2z}$ is the   $z$-projection of the total spin angular momentum  ,
 $\beta=\hbar\omega_0/kT$ is the dimensionless inverse temperature,} $\omega_0=\gamma |\vec{H}_0|$)
we obtain the following  evolution of the  density matrix  \cite{FP} (we use the standard basis of the eigenstates of the operator $I_z$: $|00\rangle$, $|01\rangle$, $|10\rangle$, $|11\rangle$):
\begin{eqnarray}\label{x1}
&&
\rho(\tau)=
e^{-iH_{MQ}\tau} \rho_0 e^{iH_{MQ}\tau} =\\\nonumber
&&
\frac{1}{2(1+\cosh\beta)}\left( \begin{matrix} \cosh\beta+\cos(D\tau)\sinh\beta & 0& 0& i\sin(D\tau)\sinh\beta\\
0 & 1& 0& 0\\
0 & 0& 1& 0\\
-i\sin(D\tau)\sinh\beta & 0& 0& \cosh\beta-\cos(D\tau)\sinh\beta \end{matrix} \right)
\end{eqnarray}
which is the solution to the Liouville equation ($\hbar =1$)
\begin{eqnarray}\label{Liouville}
i \frac{d\,\rho(\tau)}{d\,\tau} = [H_{MQ},\rho(\tau)].
\end{eqnarray}
Note, that the density  matrix (\ref{x1})  has the  structure
\begin{eqnarray}
\rho=\left(
\begin{array}{cccc}
\rho_{11}&0&0&\rho_{14}\\
0&\rho_{22}&0&0\\
0&0&\rho_{33}&0\\
-\rho_{14}&0&0&\rho_{44}
\end{array}
\right),\;\;\;\sum_{i=1}^4\rho_{ii}=1,
\end{eqnarray}
which is a particular case of the more general  X-matrix whose discord has been  analytically 
studied in \cite{ARA}. { In our case, formulas derived in \cite{ARA} may be simplified yielding explicit expression for the discord $Q$ as a function of  $\beta$ and $\tau$. }
By definition \cite{OZ,HV,ARA} the discord is a difference of two components: the total mutual information $I(\rho)$ and the classical correlations $C(\rho)$, i.e.
\begin{equation}\label{g_discord}
Q(\rho)=I(\rho)-C(\rho).
\end{equation}
Here the total  mutual information $I(\rho)$ is given by \cite{OZ}
\begin{equation}\label{I}
I(\rho)=S(\rho^A)+S(\rho^B)+\sum_{j=0}^3 \lambda_j \log_2 \lambda_j,
\end{equation}
where $A$ and $B$ mark the first and the second spins respectively, 
$\rho^A=Tr_B\rho={\mbox{diag}}(\rho_{11}+\rho_{22}, \rho_{33}+\rho_{44})$, 
$\rho^B=Tr_A\rho={\mbox{diag}}(\rho_{11}+\rho_{33}, \rho_{22}+\rho_{44})$ are the reduced density matrices and $\lambda_j$ are 
the eigenvalues of the density matrix (\ref{x1}):
\begin{equation}
\label{eigenvalues}
\begin{array} {c}
\lambda_0= \dsfrac{\cosh\beta+\sinh\beta}{2(1+\cosh\beta)},\; \lambda_1=\dsfrac{\cosh\beta-\sinh\beta}{2(1+\cosh\beta)}, \\ \\
\lambda_2=\lambda_3=\dsfrac{1}{2(1+\cosh\beta)}.
\end{array}
\end{equation}
Quantities $S(\rho^A)$ and $S(\rho^B)$ are the appropriate entropies \cite{OZ} ($S(\rho)=-{\mbox{Tr}}(\rho\log_2\rho)$)
 which read in our case 
\begin{eqnarray}\label{SpAB}
&&
S(\rho^A)=S(\rho^B)= \\\nonumber
&&
-\frac{1}{2} \log_2 \frac{(\cosh\beta+1)^2 - \xi^2\sinh^2\beta}{4(1+\cosh\beta)^2}
-\frac{\xi\sinh\beta}{2(1+\cosh\beta)}
\log_2\frac{\cosh\beta+1+\xi\sinh\beta}{\cosh\beta+1-\xi\sinh\beta}
\end{eqnarray}
where we introduce the time-dependent  parameter $\xi=|\cos(D\tau)|$, $0\le \xi \le 1$.
Following  ref.\cite{ARA}, we assume that the projective measurements are performed over the subsystem $B$. In our case, the expression for the classical correlations 
  $C(\rho)$ (see refs.\cite{OZ, ARA}) in Eq.(\ref{g_discord})  is simplified:
\begin{equation}\label{correl}
C(\rho)=S(\rho^A)-\min_{\eta=\{0,1\}} \Omega(\eta,\beta,\xi),
\end{equation}
where 
\begin{equation}
\Omega(\eta, \beta, \xi)=p_0S_0+p_1S_1
\end{equation}
with \cite{FZ}
\begin{eqnarray}\label{S}
&&S_i = -\frac{1-\theta^{(i)}}{2}\log_2\frac{1-\theta^{(i)}}{2}-
                 \frac{1+\theta^{(i)}}{2}\log_2\frac{1+\theta^{(i)}}{2},
\\\label{p}
&&p_i=\frac{1}{2} \Big(1+(-1)^i\eta(2(\rho_{11}+\rho_{33})-1) \Big),\\\label{theta}
&&\theta^{(i)}=\frac{1}{p_i}\Big[(1-\eta^2) |\rho_{14}|^2+\\\nonumber
&&
\frac{1}{4}
\Big(
2(\rho_{11}+\rho_{22})-1 +(-1)^i \eta(1-2(\rho_{22}+\rho_{33}))\Big)^2 
 \Big]^{1/2},
\;\;\;i=0,1.
\end{eqnarray}
Here  $\eta$ is an arbitrary parameter, $0\le \eta\le 1$. 
%related with the parameters  $k$, $l$ and $z_3$ used in \cite{ARA} by $\eta=k-l=z_3$.
The minimum of $\Omega(\eta, \beta, \tau)$ in eq.(\ref{correl}) corresponds to $\eta=0$,
\begin{eqnarray}\label{Om0}
\Omega(0, \beta , \xi)=
\log_2 (1+e^\beta) -\frac{\beta e^\beta}{\ln 2(1+e^\beta)},
\end{eqnarray}
which is proven in Appendix. Thus, the optimization problem, 
which is inherent in the definition of the quantum discord \cite{OZ}, is  completely solved for the dimer in the MQ NMR experiment. { Remember, that optimization problem for general X-matrix \cite{ARA} is reduced to the minimum of six values. 

Now we are able to write a single expression for the discord.} 
In fact, combining eqs.(\ref{g_discord},\ref{I},\ref{correl}) and taking into account that 
$\sum_{i=0}^3 \lambda_i \log_2\lambda_i =-2\, \Omega(0,\beta,\xi)$ (which can be simply derived substituting $\lambda_i$ from Eqs.(\ref{eigenvalues}) into 
the expression $\sum_{i=0}^3 \lambda_i \log_2\lambda_i$) 
 we obtain the following expression for the quantum discord:
\begin{eqnarray}\label{Qf}
&&
Q(\beta,\xi)=S(\rho^A) - \Omega(0,\beta,\xi)=\\\nonumber
&&
\log_2(1+e^\beta) - \frac{\beta}{\ln 2(1+e^\beta)}
-\frac{1}{2} \log_2 \Big((\cosh\beta+1)^2 - \xi^2\sinh^2\beta\Big)
-\\\nonumber
&&
\frac{\xi\sinh\beta}{2(1+\cosh\beta)}
\log_2\frac{\cosh\beta+1+\xi\sinh\beta}{\cosh\beta+1-\xi\sinh\beta}.
\end{eqnarray}
{ The three-dimensional plot of the discord $Q$ (\ref{Qf}) as a function of dimensionless inverse temperature $\beta$ and time-dependent parameter  $\xi$ is depicted in Fig.\ref{3d}

\begin{figure}
\epsfig{file=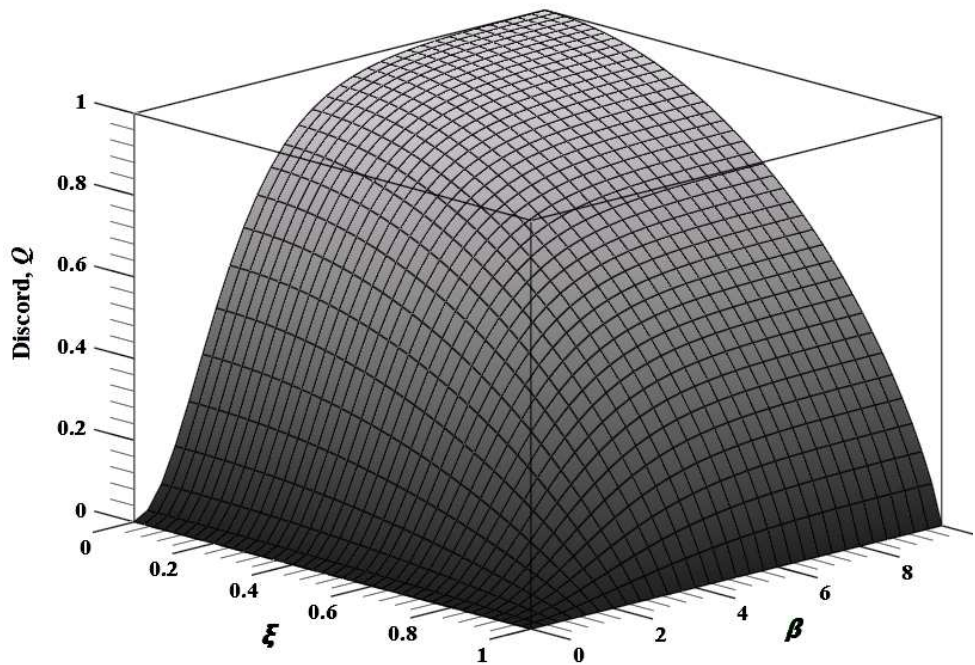, scale=0.6, angle=0} 
\caption{The discord $Q$ as a function of dimensionless inverse temperature $\beta$ and time-dependent parameter $\xi$
}
\label{3d}
\end{figure}
}
%%%%%%%%%%%%%%%%%%%
\section{Quantum discord and concurrence  as  functions of second-order coherence intensity}
\label{Section:coh}

Since the discord has been introduced as a measure of quantum information, it is important to have a  tool allowing one to measure the discord in experiments \cite{Yu}. 
Regarding the MQ NMR experiment, the MQ coherence intensities should be taken as measurable quantities for this purpose \cite{WSWP,BMGP}. 

Remember, that the standard MQ NMR experiment consists of four periods \cite{BMGP}: (i) the preparation period,  where the $H_{MQ}$ Hamiltonian is generated  by  the special  sequence  of radio-frequency pulses; duration of this period is $\tau$, (ii) the evolution period, where the system evolves without radio-frequency pulses;  the effective Hamiltonian on this period is $\Delta I_z$, where $\Delta$ is some constant parameter; duration of this period is $t$, (iii) the mixing period, where the system is irradiated by the special pulse sequence generating the
 $-H_{MQ}$ Hamiltonian; duration of this period is $\tau$, and (iv) detection. 
 { Due to the special structure of the MQ Hamiltonian  $H_{MQ}$, the transitions between eigenstates with different $z$-projections of the total spin angular momentum become possible. The longitudinal magnetization  after the mixing period  involves signal 
 generated by these transitions. The MQ NMR experiment  allows one to split the 
 longitudinal magnetization into  so-called MQ NMR coherences of different orders $k$. The $k$th-order
 MQ NMR coherence combines contributions from the transitions between states whose $z$-projections of the total spin angular momentum  differ by $k$. Such transitions are possible  in clusters of at least $k$  correlated spins which are formed due to the DDI interactions in the MQ NMR experiment.} We do not represent details of the MQ NMR experiment, which may be found, for instance, in \cite{BMGP}. We only remind
 how different MQ coherence intensities $G_k(\tau)$ may be selected from the Fourier transform of the  longitudinal magnetization $\langle I_z \rangle$ after the mixing period of the MQ NMR experiment.

The  longitudinal magnetization $\langle I_z \rangle$ after the mixing period reads \cite{G}:
\begin{eqnarray}\label{I0}
\langle I_z(\tau,t) \rangle ={\mbox{Tr}}\;( \rho(\tau,t) I_z ),
% =\sum_{k=-N}^N e^{-i k \Delta t} G_k(\tau),
\end{eqnarray} 
where  $\rho(\tau,t)$ is the density matrix after the mixing period,
\begin{eqnarray}
\rho(\tau,t)= e^{i H_{MQ} \tau}  e^{-i \Delta I_z t} e^{-i H_{MQ} \tau} 
 \rho_0
 e^{i H_{MQ} \tau}  e^{i \Delta I_z t} e^{-i H_{MQ} \tau} .
\end{eqnarray} 
Then Eq.(\ref{I0}) may be written as follows:
\begin{eqnarray}\label{I2}
\langle I_z(\tau,t) \rangle ={\mbox{Tr}}\;\left( 
 e^{-i \Delta I_z t}
\rho(\tau) e^{i \Delta I_z t} \rho^{ht} (\tau)  \right)
 =\sum_{k=-N}^N e^{-i k \Delta t} G_k(\tau),
\end{eqnarray} 
where $\rho$ is defined in Eq.(\ref{x1}), while  the density matrix  $\rho^{ht}$ reads
\begin{eqnarray}\label{rho}
\rho^{ht}(\tau) = e^{-i H_{MQ} \tau} I_z e^{i H_{MQ} \tau},
\end{eqnarray}
and $G_k$ is the $k$th-order  MQ coherence intensity,
\begin{eqnarray}\label{Gk}
G_k = \rho_k \rho^{ht}_{-k}.
\end{eqnarray}
In Eq.(\ref{Gk}),  we use the following representations of the density matrices:
\begin{eqnarray}
\rho =\sum_{k=-N}^N \rho_k,\;\;\rho^{ht} =\sum_{k=-N}^N \rho^{ht}_k,
\end{eqnarray}
where $\rho_k$ and $\rho^{ht}_k$ are those parts of the matrices $\rho$ and $\rho^{ht}$ respectively which are responsible for the $k$th-order coherence \cite{FL3}, i.e they satisfy the conditions
\begin{eqnarray}
e^{-i  \Delta I_z t} \rho_k e^{ i \Delta I_z t} = e^{-i k \Delta t} \rho_k,\;\;\;\;
e^{-i  \Delta I_z t} \rho^{ht}_k e^{ i \Delta I_z t} = e^{-i k \Delta t} \rho^{ht}_k.
\end{eqnarray}

Let us remember \cite{FL1,FL2,FL3} that only the zeroth- and $\pm$2nd-order MQ NMR coherence intensities are generated in the case of dimer (which might be considered as a particular model with the nearest neighbor interaction) and  the intensities $G_{\pm 2}$ for the state (\ref{x1})  are defined as 
  \cite{D,FM,FP}
\begin{eqnarray}\label{G0G2}
G \equiv G_{\pm 2}= \rho_{14}(\tau) \rho^{ht}_{41}(\tau)=\frac{1}{2}\tanh \frac{\beta}{2} \sin^2(D\tau)=
\frac{1}{2}\tanh \frac{\beta}{2} (1-\xi^2),
\end{eqnarray}

{ 
Now we proceed to study the relation between the discord $Q$ and  the coherence intensity $G$.
We follow ref. \cite{FP}, where the concurrence (as a measure of quantum  entanglement) has been derived as a function of $\beta$ and $\tau$,
\begin{eqnarray}\label{C}
C(\beta,\xi) ={\mbox{max}} \left(0, \frac{\sqrt{1-\xi^2}\sinh\beta-1}{2 \cosh^2\frac{\beta}{2}}\right),\;\;
\xi=|\cos(D\tau)|,
\end{eqnarray}
and its dependence on the second-order coherence intensity $G$  has been studied for dimer. }
We show that the second-order coherence intensity  is an appropriate observable quantity, which may serve to detect not only the concurrence, but also  the quantum discord in the MQ NMR experiment with dimers. 

We describe the relation between $Q$ and $G$  in  two  ways resulting to the discord as a function of either $\beta$ and $G$ (Sec.\ref{Section:betG})  or $G$ and $\xi$ (Sec.\ref{Section:Gxi}). We also  compare these relations with the analogous relations for concurrence.

%%%%%%%%%%%%%%%%%%%%%%%%%%%
 \subsection{ Discord $Q$  and concurrence $C$  as functions of dimensionless  inverse temperature $\beta$ and coherence intensity $G$} 
 \label{Section:betG}
\begin{figure}
\epsfig{file=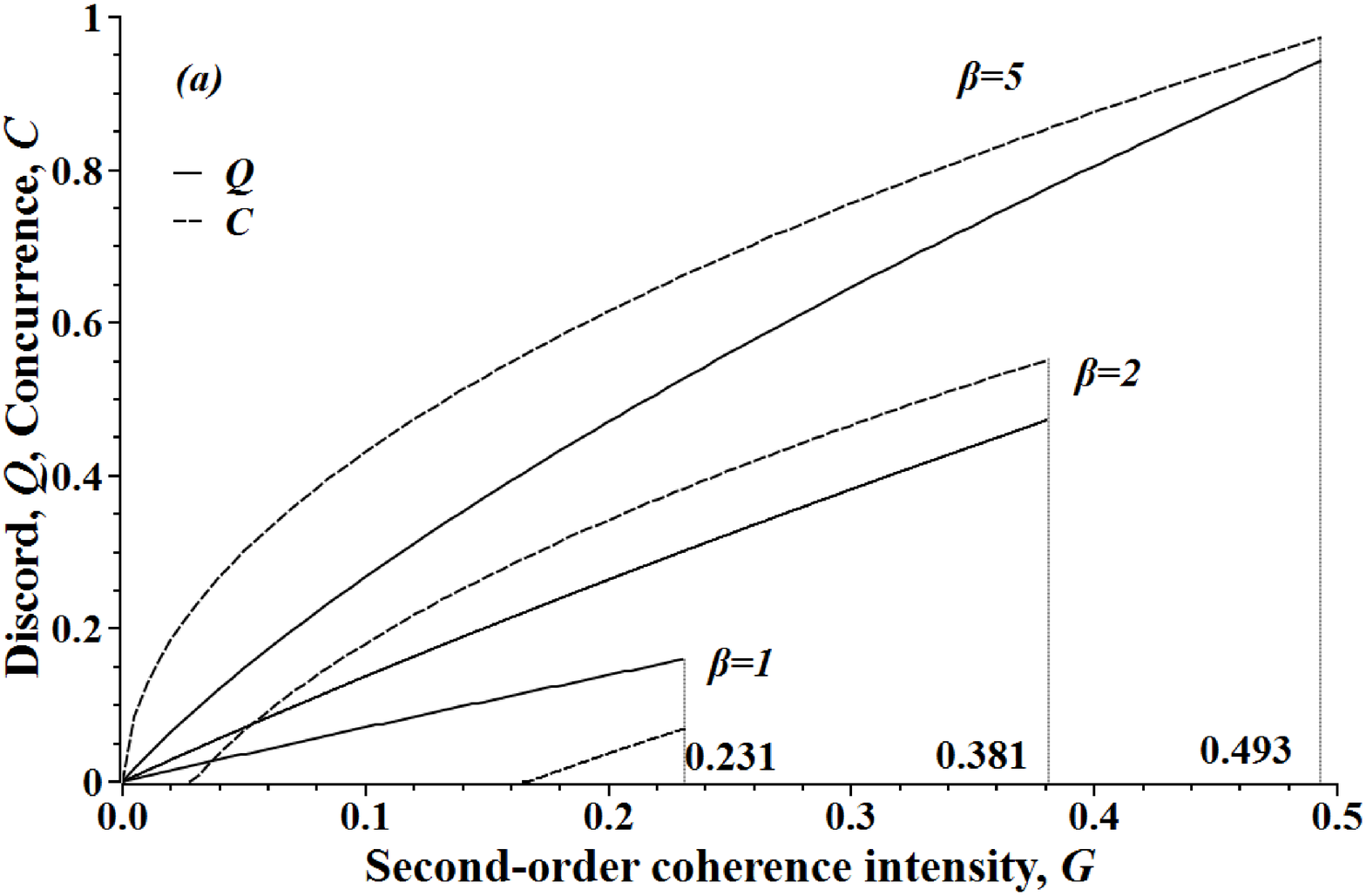, scale=0.3, angle=0} 
\epsfig{file=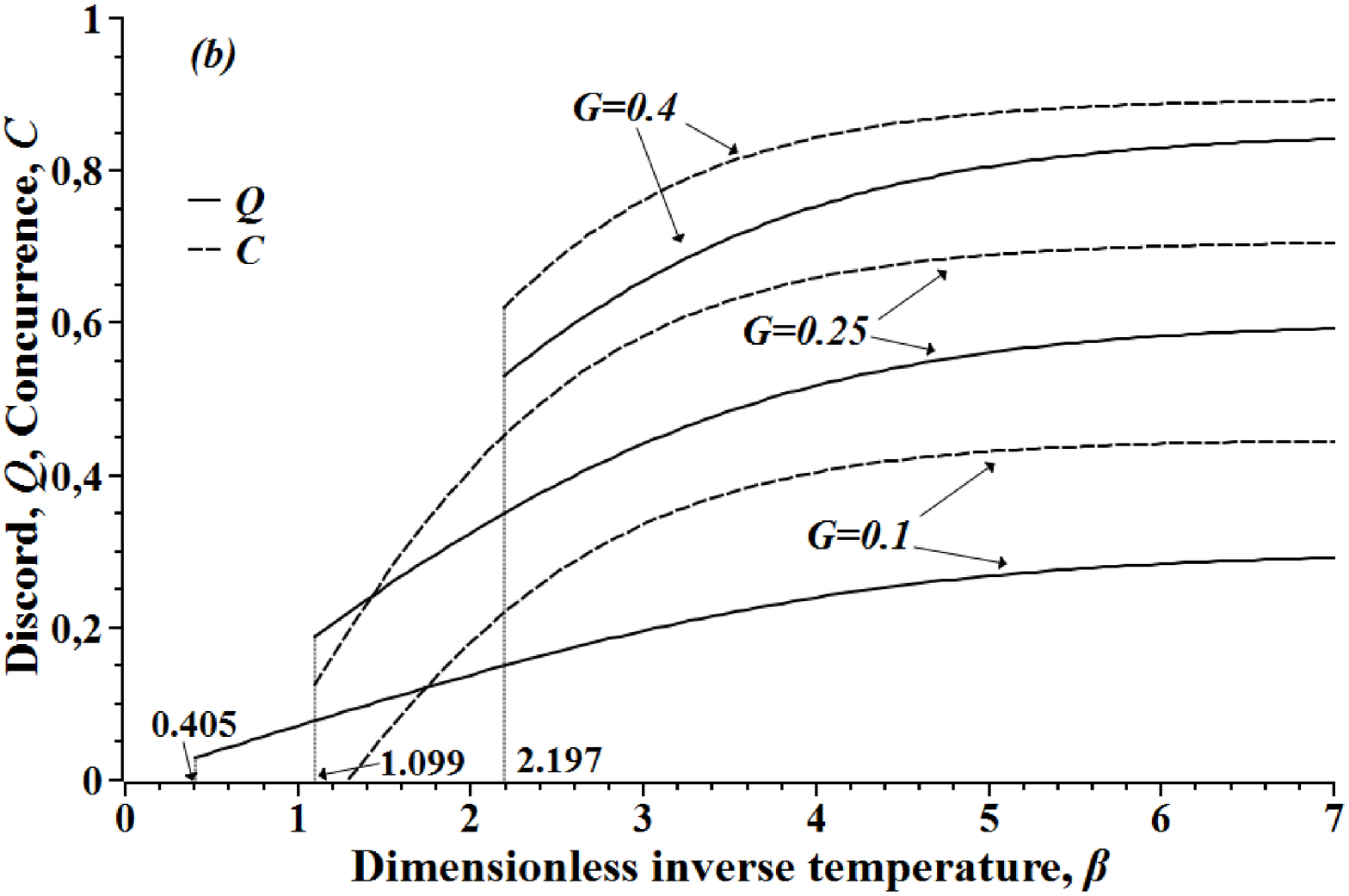, scale=0.34, angle=0}
\caption{The discord $Q$ and the concurrence $C$ as functions of the second-order coherence intensity $G$ and inverse temperature $\beta$. ($a$)  The  discord $Q$  and the concurrence $C$  
versus the second-order coherence intensity $G$ at different  $\beta$, $\beta=1, 2, 5$. The appropriate maximal values of 
$G$   are following: $G^{(1)}_{max}$=0.231, 0.381, 0.493. At these $G$, discord and concurrence reach their maximal values:
$Q_{max} = 0.160,\;0.473,\;0.942$, $C_{max}=0.069,\;0.552, 0.973$. The concurrence becomes positive at $G^{(1)}_{min}$=0.167, 0.029, $9.0\times 10^{-5}$. ($b$) The  discord $Q$  and the concurrence $C$  
versus the inverse temperature  $\beta$, at different  $G$, $G=0.1, 0.25, 0.4$. The appropriate minimal values of 
$\beta$  are following: $\beta^{(1)}_{min}$=0.405, 1.099, 2.197. At these $\beta$, discord and concurrence reach their minimal values:
$Q_{max} = 0.029,\;0.189,\;0.531$, $C_{max}=0,\;0.125, 0.620$. The  concurrence becomes positive at $\beta^{(2)}_{min}=1.295$ (for $G=0.1$) and at $\beta^{(1)}_{min} =\;1.099,\; 2.197$ (for $G=0.25,\;0.4$)
}
\label{graph2}
\end{figure}

 Using Eq.(\ref{G0G2}) we may express
 $\xi$ in terms of $G$,
\begin{eqnarray}\label{xi}
\xi =\sqrt{1-\frac{2 G}{\tanh ({\beta/2})}},
\end{eqnarray}
and substitute it into eq.(\ref{Qf}) to end up with  the discord $Q$ as the function of $\beta$ and $G$: $Q(\beta,G)$. { The explicit expression for 
$Q(G,\xi)$ is very complicated and it is not represented here.
Substituting Eq.(\ref{xi}) for $\xi$ into Eq.(\ref{C}) we obtain the concurrence as the function of $\beta$ and $G$  \cite{FP}:
\begin{eqnarray}
\label{CGG}
C(\beta,G)={\mbox{max}}\left(0,\sqrt{2 G \tanh\frac{\beta}{2}} -\frac{1}{2\cosh^2\frac{\beta}{2}}\right),
\end{eqnarray}
The concurrence $C$ and the discord  $Q$ as  
functions of $G$ at three different values of $\beta$ are compared  in Fig.\ref{graph2}a.
In this case, in accordance with Eq.(\ref{xi}),
$G$ varies  in the range
\begin{eqnarray}\label{Gint1}
0\le G \le G^{(1)}_{max}(\beta)=\frac{1}{2}\tanh ({\beta/2}),
\end{eqnarray}
so that the maximal values of the concurrence and discord are $C(\beta,G^{(1)}_{max})$ and $Q(\beta,G^{(1)}_{max})$ respectively.

According to Eq.(\ref{CGG}), the concurrence is positive (i.e. the state is entangled) if  \cite{FP}
\begin{eqnarray}
\label{CGG22}\label{G1min}
G^{(1)}_{min}(\beta)< G(\tau,\beta) < G^{(1)}_{max}(\beta),\;\;
G^{(1)}_{min}(\beta)=\frac{1}{4\sinh \beta \cosh^2\frac{\beta}{2}},
\end{eqnarray}
unlike the  discord $Q$, which  is positive over the whole interval (\ref{Gint1}).

The comparison of concurrence $C$ and discord  $Q$  as functions of $\beta$ at three different values of $G$ is depicted in   
 Fig.\ref{graph2}b. 
 Here parameter $\beta$ is restricted as
 \begin{eqnarray}\label{bet1min}
\beta^{(1)}_{min}(G)\le \beta,\;\;\;
\beta^{(1)}_{min}(G) = 2 \tanh^{-1}(2 G) ,
\end{eqnarray}
which follows from Eq.(\ref{xi}).
Thus the minimal value of discord  is
$Q(\beta^{(1)}_{min},G)$.
Again, the concurrence is positive  if 
\begin{eqnarray}\label{bet_ent}
\max(\beta^{(1)}_{min}(G),\beta^{(2)}_{min}(G))< \beta,
\end{eqnarray}
where $\beta^{(2)}_{min} =2\tanh^{-1}(X^2)$ and $X$ is the unique (for positive $G$) positive solution to the equation 
\begin{eqnarray}\label{X1}
\frac{X^4}{2} + \sqrt{2 G} \;X -\frac{1}{2}=0,
\end{eqnarray}
which is obtained from Eq.(\ref{CGG}). Thus, the minimal value of the concurrence is $C(\max(\beta^{(1)}_{min},\beta^{(2)}_{min}),G)$ which is zero if $\beta^{(1)}_{min}\le \beta^{(2)}_{min}$.

Emphasize that the discord may be valuable even if the concurrence is zero, which is confirmed by  Figs. \ref{graph2}a and \ref{graph2}b ($G=0.1$).}

%%%%%%%%%%%%%%%%
\subsection{ Discord $Q$ and concurrence $C$ as  functions of  the coherence intensity  $G$ and time-dependent parameter $\xi$}
\label{Section:Gxi}

Similarly, we may express the dimensionless inverse  temperature  $\beta$  in terms of the coherence intensity $G$ from Eq.(\ref{G0G2}),
\begin{eqnarray}\label{beta}
\beta =2\tanh^{-1}(2 G/(1-\xi^2)),
\end{eqnarray}
and substitute it into eq.(\ref{Qf}) obtaining the discord $Q$ as a function of $G$ and $\xi$: $Q(G,\xi)$. 
The explicit expression for 
$Q(G,\xi)$  is not represented here, because it is too cumbersome.

\begin{figure}
\epsfig{file=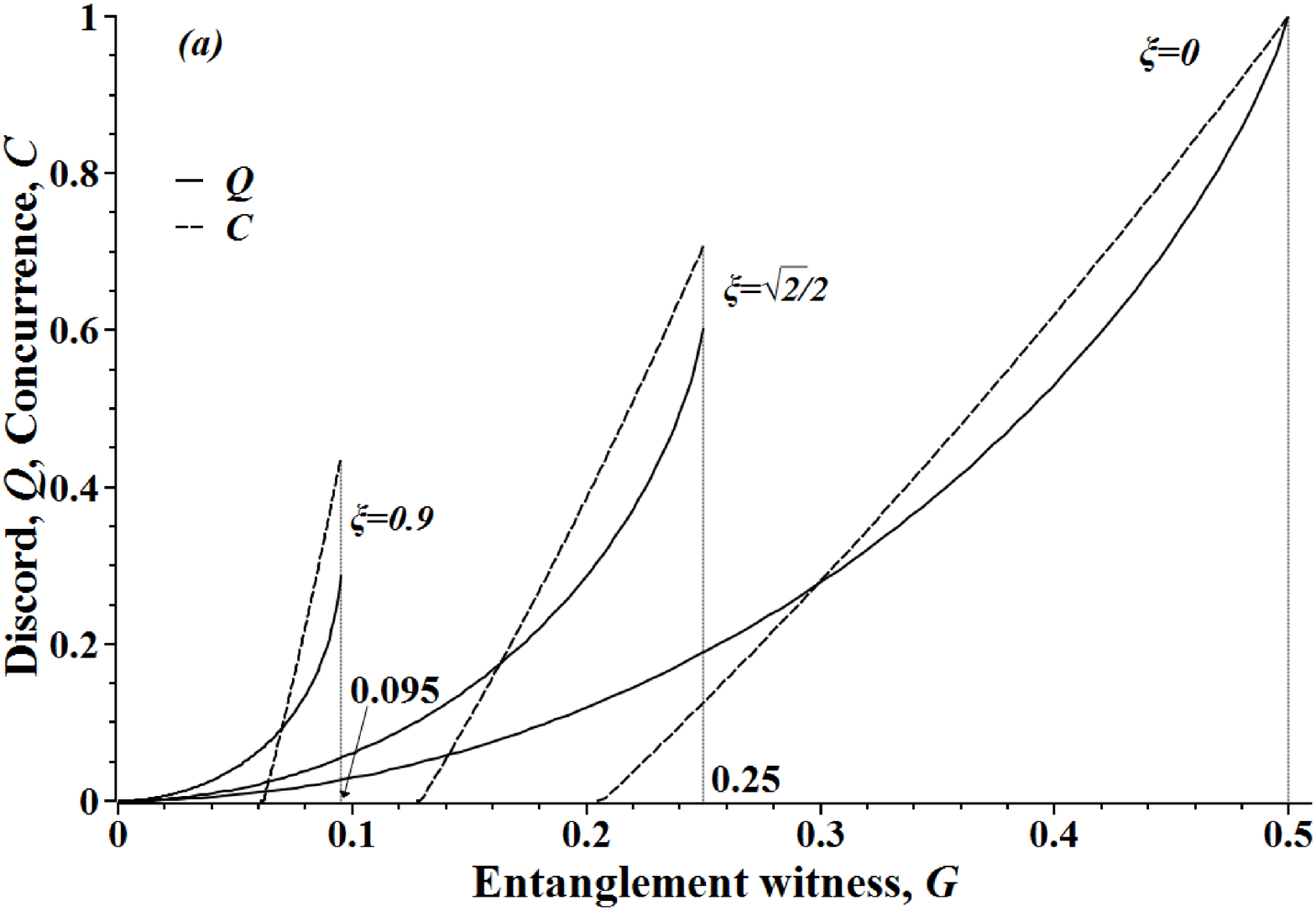, scale=0.3, angle=0}
\epsfig{file=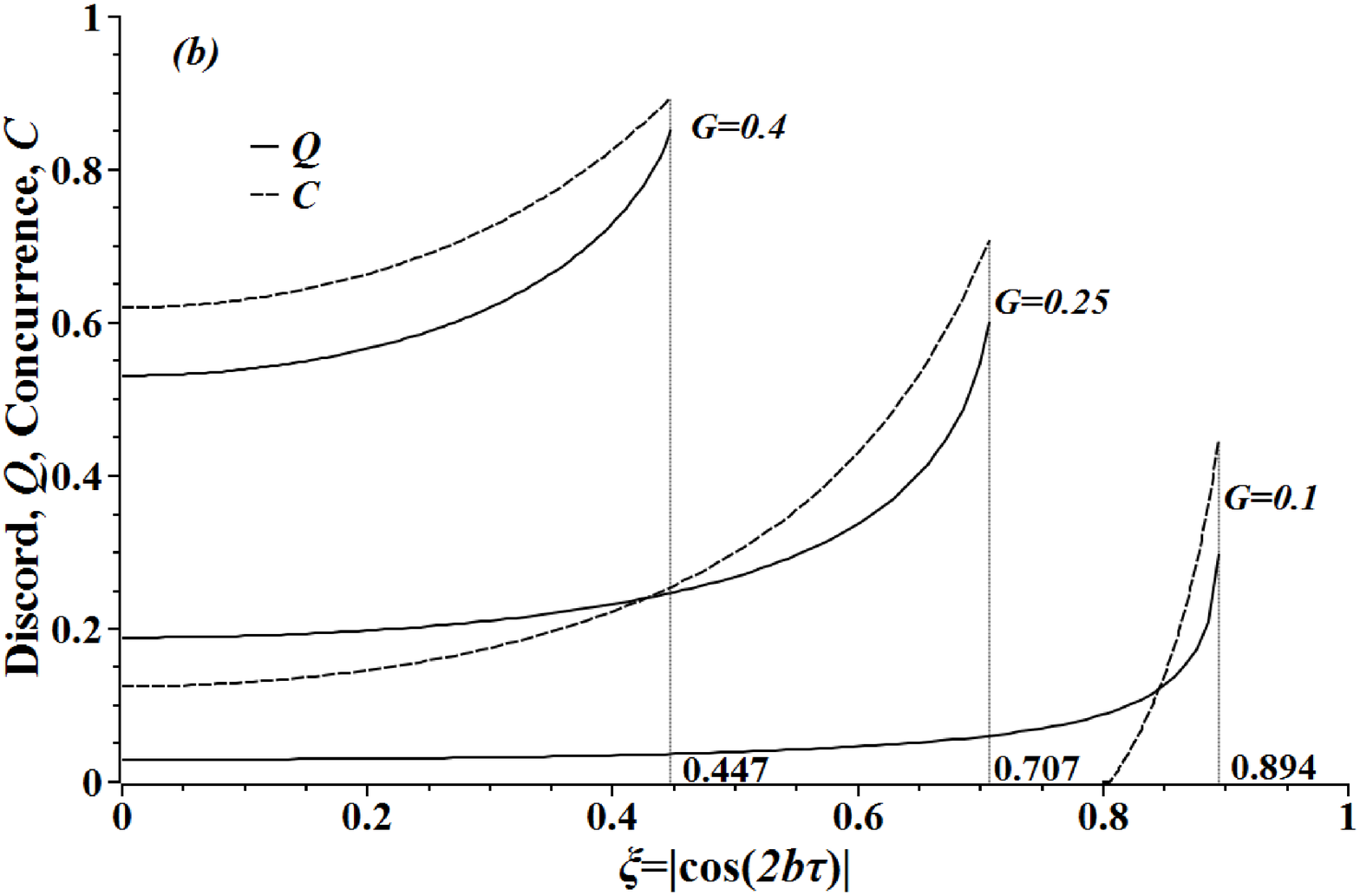, scale=0.3, angle=0} 
\caption{The discord $Q$ and concurrence $C$ as functions of the second-order coherence intensity and time-dependent parameter $\xi$. ($a$) The discord and concurrence  versus  $G$ at different  $\xi=0.9, \; \sqrt{2}/2,\;0$. The appropriate maximal values of 
$G$ are following: $G^{(2)}_{max}$=0.095, 0.250, 0.500. At these $G$, discord and concurrence reach their  maximal values:
$Q_{max} =0.286,\;0.601,\; 1$, $C_{max}=0.436,\;0.707,\;1$. The concurrence becomes positive at $G^{(2)}_{min} = 
0.062,\;0.129,\;0.207$. ($b$) The discord and concurrence  versus  $\xi$ at different  $G=0.1, \; 0.25,\;0.4$. The appropriate maximal values of 
$\xi$ are following: $\xi^{(2)}_{max}$=0.894, 0.707, 0.447. At these $\xi$, discord and concurrence reach their  maximal values:
$Q_{max} =0.298,\;0.601,\; 0.850$, $C_{max}=0.447,\;0.707,\;0.894$. The concurrence becomes positive  at $\xi^{(2)}_{min} = 0.806$ (for $G=0.1$) and at $\xi=0$ (for $G=0.25,\;0.4$)
}
\label{graph3}
\end{figure}

{
Eliminating $\beta$ from Eq.(\ref{C}) using Eq.(\ref{beta}) we obtain  the  concurrence as a function of $G$ and $\xi$:
\begin{eqnarray}
\label{CGG2}
C(G,\xi)={\mbox{max}}\left(0,\frac{2 G}{\sqrt{1-\xi^2}} +\frac{2 G^2}{(1-\xi^2)^2}-\frac{1}{2}\right).
\end{eqnarray}
The discord  and the concurrence as functions of $G$  at three different values of $\xi$ are  compared   in Fig.\ref{graph3}a. 
In this case
$G$ varies  in the range
\begin{eqnarray}\label{Gint2}\label{G2max}
0\le G < G^{(2)}_{max}(\xi),\;\;\;
G^{(2)}_{max}(\xi)=\frac{1}{2}(1-\xi^2),
\end{eqnarray}
so that the maximal values of discord and concurrence  are  $Q(G^{(2)}_{max},\xi)$ and $C(G^{(2)}_{max},\xi)$. 
The almost linear behavior of the concurrence $C$ is explained by Eq.(\ref{CGG2}), which means that $C\sim G$ for $G<1$.

According to Eq.(\ref{CGG2}), the concurrence is positive (i.e. the state is entangled) if \cite{FP} 
\begin{eqnarray}
\label{CGG23}\label{G2min}
 G^{(2)}_{min}(\xi)<G \le G^{(2)}_{max}(\xi),\;\;
G^{(2)}_{min}(\xi)=\frac{1}{2} \left((1-\xi^2) \sqrt{2-\xi^2} - (1-\xi^2)^{3/2}\right),
\end{eqnarray}
unlike the discord, which is positive over the whole  interval (\ref{Gint2}).

 Comparison of the discord and the concurrence as  functions of $\xi$  at three different values of $G$ is
 depicted in  Fig.\ref{graph3}b. Here we have (according to eq.(\ref{beta}))
 \begin{eqnarray}
 0\le \xi \le \xi^{(2)}_{max}(G) ,\;\; \xi^{(2)}_{max}(G) =\sqrt{1-2 G }.
 \end{eqnarray}
 Thus the maximal values of the discord and concurrence are $Q(G,\xi^{(2)}_{max})$ and 
 $C(G,\xi^{(2)}_{max})$ respectively.
 The concurrence is positive if 
 \begin{eqnarray}
 \xi^{(2)}_{min}(G)<\xi\le\xi^{(2)}_{max}(G) ,
 \end{eqnarray}
where $\xi^{(2)}_{min}(G) =\sqrt{1-X^2}$ and $X$ is the unique  (for positive $G$) positive solution to the equation 
\begin{eqnarray}
\frac{X^4}{2} - 2 G X^3 -2 G^2=0,
\end{eqnarray}
which is derived from Eq.(\ref{CGG2}).

Similar to Fig.\ref{graph2}, Figs. \ref{graph3}a and \ref{graph3}b ($G=0.1$)  demonstrate that the discord may be valuable even if the concurrence is zero.}

Thus, having functions  $Q(\beta,G)$ and $C(\beta,G)$  (or $Q(G,\xi)$ and $C(G,\xi)$ ) we are able to find the discord and concurrence measuring the second order coherence intensity at the given inverse temperature $\beta$ (or at the given value of the parameter  $\xi=|\cos(D\tau)|$).

{ The graphs in Figs.\ref{3d}, \ref{graph2} and \ref{graph3} demonstrate that the maximal values of both discord and concurrence equal to unity, which corresponds to $\xi=0$, $G=1/2$ (or $\beta\to\infty$, $G=1/2$).  

Finally we remark,  that the limit $\beta\to\infty$ ($T=0$)  yields the pure state.  In fact, the density matrix (\ref{x1}) reads in this limit:
\begin{eqnarray}
\label{pure}
\rho(\tau) = \frac{1}{2}\left(\begin{array}{cccc}
1+\cos(D\tau)&0&0&i \sin(D\tau)\cr
0&0&0&0\cr0&0&0&0\cr
-i \sin(D\tau)&0&0&1-\cos(D\tau)
\end{array}\right),
\end{eqnarray}
so that $\rho^2=\rho$, which is the criterion of pure state. It is known \cite{DSC} that    the discord coincides with  the entanglement for this state.
}

%%%%%%%%%%%%%%%%%%%
\section{Conclusions}
\label{Section:conclusions}
We show that the quantum discord in  dimers (whose spin dynamics is governed by the Hamiltonian $H_{MQ}$ (\ref{HMQ}) )  may be expressed in terms of the second-order coherence intensity, similar to the entanglement \cite{FP}. Due to this fact, the discord, which is the function of the dimensionless inverse temperature and time (i.e. $\beta$ and $\xi=|\cos(D\tau)|$), may be considered either as a function of the dimensionless inverse temperature $\beta$ and  the coherence intensity $G$ (i.e. $Q(\beta,G)$) or as a function of   the second-order coherence intensity $G$ and the time $\tau$ (i.e. $Q(G,\xi)$). Thus, considering the MQ NMR experiment with dimer,
one can  detect the quantum discord in the sample  measuring the second-order coherence intensity therein. 

{ Comparing the discord and the concurrence in Figs.\ref{graph2} and \ref{graph3}  we observe that there is some quantitative difference between discord and concurrence, while the general behavior of these two measures of quantum correlations is very similar. To the most important difference one must refer the fact that the concurrence may be zero for positive $G$, while discord vanishes only at $G=0$. Even if concurrence is zero, the discord may be valuable. This confirms the advantage of discord as a measure of quantum correlations in  a system. 
The fact that graphs of concurrence are above the graphs of discord in many cases (see Figs.\ref{graph2}, \ref{graph3})  indicates that these measures  take into account quantum correlations in different manners.

One has to emphasize that the requirement to have non-zero discord is much less rigorous than the requirement to have non-zero entanglement in a quantum system, which is confirmed by  Figs. \ref{graph2} and \ref{graph3}. This enriches the variety of quantum systems applicable in quantum devises. In particular, one has to note that the non-zero discord has been observed in the liquid-state NMR \cite{ACSAMSBSOS,PMTL,FMAVSO}, where the quantum entanglement is absent. This creates a new  material basis for development of 
quantum information devises.}

Authors thank Professor E.B.Fel'dman for useful discussions.
This work is supported by the Program of the Presidium of RAS No.8 ''Development of methods of obtaining chemical compounds and creation of new materials''.

\section{Appendix}
Let us show that 
\begin{eqnarray}
\label{min}
\min\Big( \Omega(0, \beta, \xi),\Omega(1, \beta, \xi)\Big)=\Omega(0, \beta, \xi).
\end{eqnarray}

The explicit expression for $\Omega(0, \beta, \xi)$ is given in Eq.(\ref{Om0}), while   $\Omega(1, \beta, \xi)$ reads:
\begin{eqnarray}\label{ff0}
\label{ff}
\Omega(1, \beta, \xi)&=&
\frac{1}{2}\log_2 ((1+\cosh \beta)^2-\xi^2\sinh^2\beta)-\\\nonumber
&& \frac{\cosh \beta}{2(1+\cosh \beta) }\log_2 (\cosh^2 \beta-\xi^2\sinh^2\beta)-\\\nonumber
&&
\frac{\xi \sinh\beta}{2(\cosh\beta+1)}\log_2 
\frac{(1+\cosh \beta-\xi\sinh\beta)(\cosh\beta+\xi\sinh\beta)}{(1+\cosh \beta+\xi\sinh\beta)(\cosh\beta-\xi\sinh\beta)}.
\end{eqnarray}
We see that $\Omega(0, \beta , \xi)$ does not depend on $\xi$ and $\Omega(1, \beta , 1)=
\Omega(0, \beta , 1)$. 
Thus, in order to prove equality (\ref{min}), it is enough to 
show that   $\Omega(1, \beta, \xi)$ is the decreasing function on the interval $0\le \xi\le 1$ 
 and consequently reaches its minimal value at $\xi=1$.

For this purpose we 
calculate the derivative of eq.(\ref{ff}) with respect to $\xi$  obtaining  the following result:
\begin{eqnarray}\label{der}
\frac{d}{d\xi}\Omega(1,\beta,\xi)= 
\frac{\sinh\beta}{2(\cosh\beta+1)}\log_2 \frac{(1+\cosh \beta)\cosh\beta-\xi^2\sinh^2\beta  - \xi\sinh\beta}{(1+\cosh \beta)\cosh\beta-\xi^2\sinh^2\beta + \xi\sinh\beta}.
\end{eqnarray}
Simple algebraic estimations show that 
\begin{equation}
0<\frac{(1+\cosh \beta)\cosh\beta-\xi^2\sinh^2\beta  - \xi\sinh\beta}{(1+\cosh \beta)\cosh\beta-\xi^2\sinh^2\beta + \xi\sinh\beta} \leqslant 1
\end{equation}
on the interval $0\le \xi\le 1$. Therefore 
the derivative (\ref{der}) is negative. 
Consequently, function (\ref{ff}) decreases on this interval and achieves the minimal value at $\xi=1$. Hence Eq.(\ref{min}) is valid.

\end{document}